\documentclass[12pt]{iopart}
\usepackage{setstack}
\usepackage{graphics}

\newcommand{\hepth}[1]{{\tt hep-th/#1}}

\newcommand{\nn}{\nonumber}

\newcommand{\ck}{{\cal K}}

\makeatletter

\@addtoreset{equation}{section}
\def\subsection{\@startsection{subsection}{2}{\z@}{-3.25ex plus
 -1ex minus -.2ex}{1.5ex plus .2ex}{\bf}}
\def\subsubsection{\@startsection{subsubsection}{3}{\z@}{-3.25ex plus%
 -1ex minus -.2ex}{1.5ex plus .2ex}{\sl}}

\begin{document}
\hfill YITP-05-77

\title[Jump-defects in non-relativistic field theories]{Jump-defects in the nonlinear Schr\"odinger model
and other non-relativistic field theories}

\author{E Corrigan$^1$ and C Zambon$^2$}

\address{$^1$ Department of Mathematics, University of York,
York YO10 5DD, U.K.}
\address{ $^2$ Yukawa Institute for Theoretical Physics,
University of Kyoto, Kyoto 606-8502, Japan}

\eads{\mailto{ec9@york.ac.uk} and
\mailto{zambon@yukawa.kyoto-u.ac.jp}}

\begin{abstract}
Recent work on purely transmitting `jump-defects' in the
sine-Gordon model and other relativistic field theories is
extended to non-relativistic models. In all the cases investigated
the defect conditions are provided by `frozen' B\"acklund
transformations and it is also shown via a Lax pair argument how
integrability will be preserved in the presence of this type of defect.
Explicit examples of the scattering of solitons by defects are
given, and
 bound states associated
with `jump-defects' in the nonlinear Schr\"odinger model are
described. Although the nonlinear Schr\"odinger model provides
the principal example, some results are also presented
for the Korteweg de Vries and modified Korteweg de Vries equations.
\end{abstract}


\section{Introduction}

The study of impurities, or defects, has a lengthy history especially in
the context of condensed matter physics (see for example the
lecture notes by Saleur \cite{Saleur98}). In the context of integrable field theories
of various types
there has been interest in studying defects both from a classical and a
quantum point of view, either theoretically or from the applications perspective.
In the quantum domain there are issues surrounding the extent to which integrability
is compatible with the reflection and transmission typical of a defect. The pioneering
work of Delfino, Mussardo and Simonetti \cite{Delf94} represents one point of
view and more recent work of Mintchev, Ragoucy and Sorba \cite{Mintchev02}
provides another. Typically,
adding a $\delta$-impurity to a classically integrable
nonlinear field theory destroys its integrability, although there
may nonetheless be interesting phenomena associated with defects of this type
 (see \cite{Goodman02} for the
behaviour of solitons in the sine-Gordon model with a $\delta$-impurity,
or \cite{Cao95} and \cite{Holmer06} for studies of some aspects of the
nonlinear Schr\"odinger model
with a $\delta$-impurity). On the other hand, there may be circumstances
where a different type of defect is able to preserve the property of classical
integrability
and it is an interesting question to investigate what those circumstances
might be. Some years ago,
 it was noticed that several types of relativistic
integrable field theory permit discontinuities (a type of defect)
 without the property of classical integrability
being destroyed
\cite{bczlandau}. Among these are free fields, Liouville theory,
the sine/sinh-Gordon model and a variety of affine Toda field
models (though possibly not all of them \cite{bcztoda}).
Besides preserving integrability, it was found that the conditions defining
the defect allowed not only energy conservation (as would be expected)
but also the conservation of a generalised momentum, including a contribution
from the defect itself (which was unexpected because of the
evident loss of translation invariance). An integrable discontinuity of
this type will be referred to as a `jump-defect' in order to distinguish
it from a (generally non-integrable) $\delta$-impurity.

The integrable jump-defects have a Lagrangian description
and their integrability is ensured by the existence of suitably adapted
Lax pairs. The specific form of the jump-defect conditions
is quite striking and makes novel use of B\"acklund
transformations which are well-known to be a bulk feature of each of
the models considered (for example, see \cite{Backlund}). With a single
jump-defect the setup is is quite straightforward and easily
described. One of the simplest examples is
provided by the sine-Gordon model, as follows.

The sine-Gordon model in the bulk \cite{Skyrme61, Scott73,Lamb} is
specified by the Lagrangian density
\begin{equation}\label{sGbulk}
    {\cal L}=\frac{1}{2}(u_t^2-
    u_x^2) - \frac{m^2}{\beta^2}
    (1-\cos\beta u)\,.
\end{equation}
A single jump-defect placed at $x=0$ is then described
by altering the Lagrangian in the following
manner, denoting the field on the left of the defect
by $u $ and the field on the right of it
by $ v $. The full Lagrangian consists of pieces from
the bulk regions ($x<0$ and $x>0$),
together with a delta function contribution at $x=0$.
Thus,  the new Lagrangian density is
given by
\begin{equation}\label{sGdefect}
{\cal L}=\theta(-x){\cal L}_u +\theta(x){\cal L}_v-
\delta(x)\left[\frac{1}{2}\left(u  v _t- v  u _t\right) +{\cal B}(u , v )\right],
\end{equation}
with
\begin{equation}\label{sGdefectpotential}
    {\cal B}=\frac{2m\sigma}{\beta^2}\cos\beta\left(\frac{u + v }{2}\right)+
    \frac{2m}{\sigma\beta^2}\cos\beta
    \left(\frac{u - v }{2}\right).
\end{equation}
The form of the additional piece is required by integrability and
leads to defect conditions linking the two fields $u$ and $v$ and
their derivatives at $x=0$, the position of the defect. The usual
variational principle reveals that the defect conditions
constitute a B\"acklund transformation `frozen' at
the defect. There are many interesting features of these defects,
both from a classical field theory point of view - especially with
regard to the behaviour of solitons - and within the quantum
context \cite{Bow05}. There is no obstacle to having several
defects at different locations (with the same or different
parameters). Moreover, they can move and scatter amongst
themselves \cite{Bow05}. As mentioned already, a novel feature exhibited by these
defects is the manner in which they may exchange both energy and
momentum with the fields on either side of the defect location.
There is nothing surprising about the energy since time
translation is unbroken, but the existence of a conserved - though
modified - momentum is quite surprising since translation
invariance is certainly lost. In fact, the defect potential
(\ref{sGdefectpotential}) may be regarded as being determined by
demanding that it be possible to find a conserved generalised
momentum functional. The ability to exchange momentum and energy
with the fields on either side of it is a defining feature of
a jump-defect, and will be used as a tool later in this paper.

As a further example, and to make clear the distinction between a jump-defect
and a $\delta$-impurity, consider two free-field situations. The jump-defect
can be described as a small field approximation to (\ref{sGdefect},\
\ref{sGdefectpotential}).
In other words,
\begin{equation}
{\cal B}(u,v)= -\frac{m\sigma}{4}(u+v)^2 -\frac{m}{4\sigma}(u-v)^2
\end{equation}
with
\begin{equation}
{\cal L}_u=\frac{1}{2}(u_t^2-u_x^2-m^2u^2), \qquad {\cal L}_v=\frac{1}{2}(v_t^2-v_x^2-m^2v^2).
\end{equation}
On the other hand, in the same notation, a $\delta$-impurity is described by the
quadratic Lagrangian,
\begin{equation}\label{deltaimpurity}
{\cal L}_\delta=\theta(-x){\cal L}_u +\theta(x){\cal L}_v-
\delta(x)\frac{1}{2}\left[\sigma uv -(u_x+v_x)(u - v )\right],
\end{equation}
leading to free Klein-Gordon equations in the bulk together with the
defect conditions
\begin{equation}
u=v, \qquad v_x-u_x=\sigma u, \qquad x=0.
\end{equation}
It is not difficult to check by explicit calculation that the former permits
a conserved total momentum functional, including a defect contribution, while
the latter does not. There is no requirement for the linear jump-defect to
satisfy $u=v$ at $x=0$, and in general there will be a discontinuity. It is easy to check
that the linear jump-defect is purely transmitting, while the $\delta$-impurity
both transmits and reflects. The jump-defect does not possess a classical bound state
but the $\delta$-impurity does for a suitable range of $\sigma$.

Besides the sine/sinh-Gordon model there are other integrable
equations which have arisen naturally in special physical systems.
Most of these are non-relativistic but might nevertheless allow
discontinuities related to jump-defect conditions of frozen-B\"acklund
type. One such example is the nonlinear (cubic) Schr\"odinger
equation (NLS); others are the Korteweg-de Vries (KdV) and
modified KdV (mKdV) equations.

As already mentioned briefly, defects in the context of integrable
field theories have been discussed before in the quantum domain, starting with
Delfino, Simonetti and Mussardo \cite{Delf94}, and elaborated subsequently by
others, including Konik and LeClair \cite{Konik97}, and Castro-Alvaredo, Fring
and G\"ohmann \cite{Fring02}.
As a consequence of their work it appeared that a defect could not allow
simultaneous reflection and transmission while maintaining integrability
(in the sense that the reflection and transmission matrices together with the
bulk S-matrix should
satisfy a set of algebraic compatibility requirements) unless
the bulk scattering matrix was independent of rapidity. On the other hand,
the sine-Gordon jump-defect is purely transmitting as far as its behaviour with
respect to solitons
is concerned and evidence was gathered in
\cite{Bow05} to support the idea that the pure transmission matrix discovered
by Konik and LeClair really describes the quantum version of the sine-Gordon
jump-defect.
Recently,
Mintchev, Ragoucy and Sorba have suggested alternative compatibility
relations that might be satisfied between the transmission and
reflection factors and the bulk S-matrix \cite{Mintchev02}. Using this
 framework  $\delta$-type impurities have been explored
in the context of NLS
\cite{Caudrelier04}.

The purpose of this paper is to demonstrate that
jump-defects can be incorporated into the NLS model
very naturally without spoiling integrability. Following an
analysis of a few of the conserved quantities - actually
sufficient to determine the form of the defect conditions - an
argument based on a generalised Lax pair is given in  sufficient
detail to demonstrate how integrability is preserved.  Similar but
less complete arguments are also given in the context of KdV and
mKdV. In all cases, remarks are made concerning the behaviour of
solitons as they encounter a jump-defect. To date, the only
genuine surprise relative to the sine-Gordon case is provided by
the strange (and still mysterious) behaviour of a `fast' soliton
in KdV. It remains to be seen what the quantum version
of the NLS jump-defect will turn out to be and how it will relate to
earlier work, if at all.

\section{The nonlinear Schr\"odinger equation}

The nonlinear Schr\"odinger equation with a cubic interaction term
will be taken to be defined by the field equation \cite{Scott73}
\begin{equation}\label{NLS}
    \rmi u_t + u_{xx} +2 u(\bar{u} u)=0\, .
\end{equation}
This may be derived in the bulk using an action
principle based on the Lagrangian
\begin{equation}\label{NLSLagrangian}
    {\cal L}=\frac{\rmi}{2}\left(\bar u u_t - \bar u_t u\right)-|u_x|^2
    + \sigma^2 |u|^4.
\end{equation}
A (real, positive) coupling constant $\sigma^2$ has been added but
then scaled away in the field equation by redefining $u$.
Henceforth it will be ignored. The sign of the cubic term is
important for some considerations. For example, the sign chosen in
(\ref{NLS}) is appropriate for a model possessing soliton
solutions (corresponding to an `attractive' interaction).

If there is a defect at $x=0$ then the bulk fields to either side
of it will be denoted $u$ and $v$, and a boundary contribution ${\bf B}$,
presumably depending on $u,v,u_t$ and $v_t$ (and possibly spatial derivatives),
will need to be added.
In other words, the full
action will be
\begin{equation}\label{NLSaction}
    {\cal A}=\int \rmd t\left[\int_{-\infty}^0 \rmd x\, {\cal L}({\bf u}) +{\bf B}+
    \int_0^\infty \rmd x\, {\cal L}({\bf v})\right].
\end{equation}
The corresponding defect conditions at $x=0$ are:
\begin{eqnarray}\label{NLSdefectconditions}
  u_x&=&\ \ \ \ \frac{\partial{\bf B}}{\partial \bar u} -\frac{\partial}{\partial t}
  \frac{\partial{\bf B}}{\partial \bar u_t }, \qquad v_x=
  -\frac{\partial{\bf B}}{\partial \bar v} -\frac{\partial}{\partial t}
  \frac{\partial{\bf B}}{\partial \bar v_t}\, ,
\end{eqnarray}
with similar expressions for the conjugate fields. Note that (\ref{NLSdefectconditions})
would need to be modified if the defect part of the action depended additionally upon the
spatial derivatives; for the time being it will be assumed these are absent.

The analogue of energy for NLS is the density
\begin{equation}\label{NLSenergy}
{\cal E}=|u_x|^2 -|u|^4,
\end{equation}
and, since the system remains time-translation invariant despite
adding the defect, the total energy including a defect
contribution will be conserved. In the bulk, as a consequence of
space-translation invariance, the momentum has the density
\begin{equation}\label{}
{\cal P}=\rmi\left(\bar u u_x- \bar u_x u\right)
\end{equation}
and it is certainly conserved. However, when there is an impurity the
momentum is not expected to be conserved since translation
invariance is broken. However, as  has been noted before in other
cases, for example in the sine-Gordon model, this is not
necessarily so. Adding a jump-defect does not spoil momentum
conservation since this kind of defect can exchange both energy
and momentum with the fields $u$ and $v$ to either side of it. To
demonstrate this, note that the total contribution from the fields
$u$ and $v$ to the momentum density is given by
\begin{equation}\label{}
    P=P(u)+P(v)
    =\int_{-\infty}^0 \rmd x\, \rmi\left(\bar u u_x- \bar u_x u\right) +\int_0^\infty \rmd x\,
    \rmi\left(\bar v v_x- \bar v_x v\right)
\end{equation}
and it easy to see that the time derivative of $P$ will depend
critically on the defect conditions since
\begin{equation}\label{Pt}
\fl P_t=\left(2|u|^4  +2\bar u_x u_x + i(\bar u u_t-\bar u_t
u)\right)_{x=0}-\left(2|v|^4  +2\bar v_x v_x+i(\bar v
v_t-\bar v_t v)\right)_
  {x=0}.
\end{equation}
To allow conservation of this charge requires that the right hand
side of (\ref{Pt}) should (if possible) be a total time derivative
of a functional of the fields evaluated at the jump-defect.
Put alternatively, if the momentum, suitably modified, is to be
preserved then the defect conditions ought to be chosen to ensure
it. This is not quite straightforward to achieve (and the conditions
associated with a $\delta$-impurity do not have this property) but, bearing in
mind (\ref{NLSdefectconditions}) the following suggestion
works perfectly. It will be seen later that it also fits more generally with the
idea of integrability. It is
sufficient to take
\begin{equation}\label{NLSB}
    {\bf B}=\Omega\, \left[\frac{\rmi}{2}\,\frac{\partial}{\partial t}\, \ln
    \left(\frac{u-v}{\bar u -\bar v}\right) +{\cal B}\right],\qquad \Omega =
    \left(\alpha^2 -  |u-v|^2\right)^{1/2}\, ,
\end{equation}
where
\begin{equation}\label{NLScalB}
    {\cal B}=\frac{1}{3}\left(\alpha^2-   |u-v|^2\right)+  \left(|u|^2+|v|^2\right),
\end{equation}
and $\alpha$ is a real parameter. Although this is  not a
completely transparent choice,  it is based on knowledge of the
NLS B\"acklund transformation and experience with the sine-Gordon equation.
It will become clear that
(\ref{NLSB}) has nice properties. With the choice (\ref{NLSB},\
\ref{NLScalB}) the defect conditions (\ref{NLSdefectconditions}) at $x=0$
 become:
\begin{eqnarray}\label{LNSconditions}
  u_x&=&-\frac{1}{2}\left[ \frac{\rmi\left(u_t-v_t\right)}{\Omega} -
  (u+v)\Omega +\frac{(u-v)(|u|^2+|v|^2)}{\Omega}\right]\nonumber\\
   v_x&=&-\frac{1}{2}\left[\frac{\rmi\left(u_t-v_t\right)}{\Omega} +
   (u+v)\Omega +\frac{(u-v)(|u|^2+|v|^2)}{\Omega}\right],
\end{eqnarray}
and the time derivative of the momentum simplifies to:
\begin{equation}\label{Qdot}
 P_t=\rmi\frac{\partial}{\partial t}\left(\bar u v - \bar v u\right)_{x=0},
\end{equation}
and thus the quite attractive combination
\begin{equation}
P-\rmi(\bar u v -\bar v u)_{x=0}
\end{equation}
is conserved. The requirement that the time-derivative of the momentum should
turn out to be expressible as a functional of the fields evaluated
at the defect is actually very strong, and finding an expression
which satisfies (\ref{Pt}) severely limits the choice of defect
condition.

On the other hand, as mentioned above, the total energy should be
conserved whatever the defect condition is, provided it does not
violate time translation invariance. Indeed, the total energy
satisfies
\begin{equation}\label{}
    E_t = \frac{\partial}{\partial t}(\Omega{\cal B})_{x=0}\, ,
\end{equation}
implying that $E-(\Omega{\cal B})_{x=0}$ is conserved.
 Notice that the
discontinuity $[v-u]_{x=0}$ at the jump-defect should not be too severe. In
fact
\begin{equation}\label{}
\nn    |u-v|_{x=0}^2\, \le\, \alpha^2,
\end{equation}
otherwise the energy would fail to be real and simply
lose its meaning. In the limit $\alpha\rightarrow 0$, the discontinuity disappears.
In this sense, the parameter $\alpha$ controls the height of the jump.

There is also a `probability', or `number', density ${\cal N}=\bar
u u$ which satisfies
\begin{equation}\label{NLSmomentumrelation}
    {\cal N}_t=\rmi\left(\bar u_x u -\bar u u_x\right)_x.
\end{equation}
Before checking that the NLS equation remains integrable even
after adding a jump-defect it is also worth examining this charge
in some detail. On the whole line this `number' is certainly
conserved (assuming suitably decaying $u$ at $x=\pm\infty$, or
periodic boundary conditions) and it is a consequence (via Noether's
theorem) of the continuous $U(1)$ symmetry of the action under the
constant change of phase
\begin{equation}\label{phasesymmetry}
    u\rightarrow \rme^{\rmi\Lambda}\,u \qquad \Lambda_x=0=\Lambda_t.
\end{equation}
When a jump-defect is added, it is natural to define,
\begin{equation}
N=\int_{-\infty}^0 \, \rmd x\, {\cal N}(u) + \int_0^\infty \, \rmd
x\, {\cal N}(v)\, ,
\end{equation}
the defect condition (\ref{LNSconditions}) leads to
\begin{equation}\label{probderv}
  N_t = \rmi\left(\bar u_x u -\bar u u_x\right)_{x=0} -
  \rmi\left(\bar v_x v -\bar v v_x\right)_
  {x=0}= \left. \frac{\partial \Omega}{\partial t}\right|_{x=0}\, .
\end{equation}
It is clear from this that the combination
\begin{equation}\label{}
    N +\Omega = N(u)+N(v) -
    \left.\left(\alpha^2 -  |u-v|^2\right)^{1/2}\right|_{x=0}
\end{equation}
is conserved. Notice again that the charge makes sense provided the defect is
not too severe with $|u-v|_{x=0}<\alpha$.

Notice that adding and subtracting the two defect conditions
(\ref{LNSconditions}) gives the following pair of relations at
$x=0$:
\begin{eqnarray}\label{NLSBacklund}
 &u_t-v_t\ =&\rmi\left( u_x+v_x\right)\Omega + \rmi(u-v)\left(|u|^2+|v|^2\right) \nonumber \\
 &u_x-v_x\ =&(u+v)\, \Omega\, .
\end{eqnarray}
If a pair of conditions such as these held for every $x$, and not merely at
$x=0$, (\ref{NLSBacklund}) would be recognised as a B\"acklund
transformation: differentiating the second
equation with respect to $x$, using the second equation to
substitute for $u_x-v_x$ wherever it appears, and adding ${\rm i}$ times
the first equation gives
\begin{equation}\label{}
    \rmi u_t+u_{xx}+2u|u|^2=\rmi v_t+v_{xx}+2v|v|^2.
\end{equation}
On the other hand, differentiating the first with respect to $x$ and the second
with respect to $t$ and subtracting leads to
\begin{equation}\label{}
    \rmi u_t+u_{xx}+2u|u|^2=-(\rmi v_t+v_{xx}+2v|v|^2).
\end{equation}
Hence, combining these manipulations, both $u$ and $v$ satisfy the
NLS equation. In fact, this is exactly the B\"acklund transformation for
NLS given  by Lamb \cite{Lamb74} (see also
\cite{Chen74},\ \cite{Konno75}).

\section{Argument for integrability}

In order to verify that introducing a jump-defect does not destroy
integrability it is necessary either to investigate additional
conserved charges or, more effciently, to examine a suitably
modified Lax pair. The latter approach is generally superior since
the Lax pair provides a generating function (as a Laurent series
in the spectral parameter) for an infinite set of independent
conserved quantities.

A Lax pair for the NLS has been provided in \cite{Zak72} (see also
\cite{FadTak}). For the above choice of conventions, a
satisfactory Lax pair is:
\begin{eqnarray}\label{nonlinSLp}
  L(u)&=&\rmi\left(\bar{u} \sigma_{+}+u \sigma_{-}\right)+\lambda \sigma_{3}\nonumber \\
M(u)&=&\rmi(2\bar{u}u+\lambda^2)\,\sigma_{3}+(\bar{u}_x-\lambda\bar{u})\sigma_{+}-
(u_x+\lambda u)\,\sigma_{-}\,,
\end{eqnarray}
where
\begin{equation}\label{Paulim}
    \sigma_+=\left(%
\begin{array}{cc}
  0 & 1 \\
  0 & 0 \\
\end{array}%
\right) \qquad
\sigma_-=\left(%
\begin{array}{cc}
  0 & 0 \\
  1 & 0 \\
\end{array}%
\right)\qquad
\sigma_3=\frac{1}{2}\left(%
\begin{array}{cc}
  1 & \phantom{-}0 \\
  0 & -1\\
\end{array}%
\right),
\end{equation}
with the property
\begin{equation}\label{zerocurvature}
    \partial_t L-\partial_x M+[M,L]=0\quad
    \Longleftrightarrow \quad \rmi u_t+u_{xx}+2u|u|^2=0\, .
\end{equation}
Moreover, (\ref{zerocurvature}) holds independently of the choice of spectral parameter $\lambda$.

Following  the ideas introduced in \cite{bcztoda} and \cite{bcdr},
modified Lax pairs can be devised which will
 build in automatically both the bulk equations and the defect conditions. In
 designing these it is
natural to introduce two extra points
  $a<0$ and $b>0$ which are the endpoints of two regions overlapping the defect,
 one on the left $R^{-}$, $-\infty <x < b$, and one on the right $R^{+}$, $a< x<\infty$
 Then, a
 suitable pair can be defined as follows:
\begin{eqnarray}\label{modLpNLSu}
\fl \hat{L}^{-}&=&L(u)\, \theta(a-x)\nn\\
\fl \hat{M}^{-}&=&M(u)+\theta(x-a)\left\{\left[u_x+\frac{\rmi}{2\Omega}(u_t-v_t)-
\frac{\Omega}{2}(u+v)+\frac{(u-v)}{2\Omega}(|u|^2+|v|^2)\right]\sigma_{-}\right.\nn\\
\fl &&\left.\ \ \
-\left[\bar{u}_x-\frac{\rmi}{2\Omega}(\bar{u}_t-\bar{v}_t)-
\frac{\Omega}{2}(\bar{u}+\bar{v})+\frac{(\bar{u}-\bar{v})}{2\Omega}(|u|^2+|v|^2)\right]
\sigma_{+}\right\},
\end{eqnarray}
 and
\begin{eqnarray}\label{modLpNLSv}
\fl \hat{L}^{+}&=&L(v)\, \theta(x-b)\nn\\
\fl \hat{M}^{+}&=&M(v)+\theta(b-x)\left\{\left[v_x+\frac{\rmi}{2\Omega}(u_t-v_t)+
\frac{\Omega}{2}(u+v)+\frac{(u-v)}{2\Omega}(|u|^2+|v|^2)\right]\sigma_{-}\right.\nn\\
\fl &&\left. \ \ \
-\left[\bar{v}_x-\frac{\rmi}{2\Omega}(\bar{u}_t-\bar{v}_t)+
\frac{\Omega}{2}(\bar{u}+\bar{v})+\frac{(\bar{u}-\bar{v})}{2\Omega}(|u|^2+|v|^2)\right]
\sigma_{+}\right\}.
\end{eqnarray}
These modified Lax pairs provide the equations of motion together
with all  the defect relations as a consequence of zero curvature
conditions of the type (\ref{zerocurvature}). In the overlapping
interval $a<x<b$, the two matrices $\hat{M}^{+}$ and $\hat{M}^{-}$
must be $x$-independent in order to maintain the zero curvature
condition, though not necessarily equal. Rather, they must be
related by the following `gauge' transformation (see \cite{bcdr}
for more details),
\begin{equation}\label{gaugetransformation}
\partial_t\ck=\ck \hat{M}^{+}(b,t)-\hat{M}^{-}(a,t)\ck\, .
\end{equation}
Note that since $\hat{M}^{\pm}$ are $x$-independent this implies
that the field $u$ and $v$ are also $x$-independent in the
overlapping interval. It can be verified that the following choice
for ${\cal K}$
\begin{equation}\label{NLSk}
\ck=I+ \frac{1}{\lambda}\left(%
\begin{array}{cc}
  \Omega & -\rmi(\bar u -\bar v) \\
  \rmi(u-v) & -\Omega\\
\end{array}%
\right),
\end{equation}
where $I$ is the $2\times 2$ identity matrix, works perfectly, in
the sense that (\ref{gaugetransformation}) is identically
satisfied. In the limit $\alpha\rightarrow 0$ the discontinuity
tends to zero and ${\cal K}\rightarrow I$, as one would expect.

If one was to consider instead the conditions associated with a $\delta$-impurity,
namely,
\begin{equation}
u_x= v_x -\sigma v,\quad v_x=u_x+\sigma u,\qquad x=0,
\end{equation}
a similar construction would lead to the conclusion that there was
no suitable matrix ${\cal K}$ and hence no Lax pair for this type of defect.
This suggests the $\delta$-impurity is not integrable, a fact apparently consistent
with numerical studies \cite{Cao95}, \cite{Holmer06}. One could envisage  more general
linear defect conditions, but the conclusion remains the same.

Once ${\cal K}$ is given, and assuming suitably decaying fields at
$\pm\infty$, or periodic boundary conditions, conserved quantities
will be generated by
\begin{equation}\label{Qgenerator}
    {\cal Q}(\lambda)={\rm Tr}\left[P\exp\left(\int_{-\infty}^a dx\, L(u)\right) \, {\cal K}
    \, P\exp\left(\int_b^\infty dx\, L(v)\right)\right].
\end{equation}
The final and necessary step would be to demonstrate that the
charges generated by (\ref{Qgenerator}) are independent and in
involution. However, the discussion of this topic will be
postponed. In the meantime,  following Ablowitz et al.
\cite{Ablowitz73}, it will be shown how the Lax pair can be used
to generate conserved charges using the integrability constraints
represented by the zero curvature condition.

The bulk Lax pair for NLS can be written alternatively as follows
\begin{equation}
L=\left(%
\begin{array}{cc}
  \lambda/2 & \rmi\bar{u} \\
  \rmi u & -\lambda/2\\
\end{array}%
\right),\qquad
M=\left(%
\begin{array}{cc}
  A & \phantom{-}B \\
 C & -A\\
\end{array}%
\right),
\end{equation}
where the entries of the matrix $M$ can be obtained from
(\ref{nonlinSLp}). Applying the zero curvature condition
(\ref{zerocurvature}) to this Lax pair, the following relations
are obtained
\begin{eqnarray}\label{integrabilityconditions}
\fl A_x&=&\rmi u B-\rmi\bar{u}C, \qquad\lambda B=2\rmi \bar{u}A-B_x+\rmi\bar{u}_t,
\qquad\lambda C=2\rmi u A+C_x-\rmi u_t\,.
\end{eqnarray}
These allow the conserved charges to be calculated as the
coefficients in an expansion in $1/\lambda$. To see how the
argument proceeds, take a couple of steps in detail. Substituting
the last two equations of (\ref{integrabilityconditions}) into the
first of (\ref{integrabilityconditions}) it follows that
\begin{equation}\label{order1overl}
A_x=-\frac{1}{\lambda}\left[(\bar{u}u)_t+\rmi(uB+\bar{u}
C)_x-\rmi(u_x B+\bar{u}_x C)\right].
\end{equation}
Repeating the substitution from (\ref{integrabilityconditions})
one finds
\begin{eqnarray}\label{order1overlsquare}
\fl A_x&=&-\frac{1}{\lambda}\left[(\bar{u}u)_t+\rmi(uB+\bar{u}
C)_x\right]-\frac{1}{\lambda^2}\left[\frac{1}{2}(\bar{u}u_x-\bar{u}_xu)_t
-\frac{1}{2}(\bar{u}u_t-\bar{u}_tu)_x\right]\nn\\
\fl &&\ \ \ \ -\frac{1}{\lambda^2}\left[(2u\bar{u}A+\rmi u_x
B-\rmi\bar{u}_xC)_x-(2\rmi u^2\bar{u}+\rmi
u_{xx})B+(2\rmi\bar{u}^2u+\rmi\bar{u}_{xx})C\right].
\end{eqnarray}
The process can be repeated to obtain further elements of the
expansion in higher order in $1/\lambda$. However,
(\ref{order1overlsquare}) is enough to illustrate the point. If
total $x$-derivatives integrate to zero (assuming $u$ and its
derivatives vanish as $|x|\rightarrow\infty$) then integrating
(\ref{order1overlsquare}) over $-\infty<x<\infty$ and
concentrating on the first two terms  gives
\begin{equation}
\frac{\partial}{\partial t}\int^{\infty}_{-\infty}\rmd
x\,(\bar{u}u)=0,\qquad \frac{\partial}{\partial
t}\int^{\infty}_{-\infty}\rmd
x\,\frac{1}{2}(\bar{u}u_x-\bar{u}_xu)=0.
\end{equation}
These are the `probability or number' and momentum
densities, respectively, used previously.

With a jump-defect the situation is different and it becomes
necessary to deal with the modified Lax pairs defined in two
regions whose overlap contains the defect. In the matrix form, the
Lax pair to use reads as follows
\begin{eqnarray}
\hat{L}^{-}&=\left(%
\begin{array}{cc}
  {\lambda/2} & \rmi\bar{u} \\
  \rmi u & -{\lambda/2}\\
\end{array}%
\right)\theta(a-x),\quad
\hat{M}^{-}=\left(%
\begin{array}{cc}
  \hat{A}^{-} & \phantom{-}\hat{B}^{-} \\
 \hat{C}^{-} & -\hat{A}^{-}\\
\end{array}%
\right)\theta(b-x)\label{LpNLSu}\\
\hat{L}^{+}&=\left(%
\begin{array}{cc}
  {\lambda/2} & \rmi\bar{v} \\
  \rmi v & -{\lambda/2}\\
\end{array}%
\right)\theta(x-b),\quad
\hat{M}^{+}=\left(%
\begin{array}{cc}
  \hat{A}^{+} & \phantom{-}\hat{B}^{+} \\
 \hat{C}^{+} & -\hat{A}^{+}\\
\end{array}%
\right)\theta(x-a),\label{LpNLSv}
\end{eqnarray}
where the entries of the matrices $\hat{M}^{+}$ and $\hat{M}^{+}$
will not be specified here but they can be obtained from the
formulae (\ref{modLpNLSu}) and (\ref{modLpNLSv}). The Lax pairs
(\ref{LpNLSu}) and (\ref{LpNLSv}) are defined in the regions
$R^{-}$ and $R^{+}$, respectively. Within the overlap the zero
curvature condition requires
\begin{equation}
\partial_x\hat{M}^{-}=\partial_x\hat{M}^{+}=0,\end{equation}  and therefore
$M^{(\pm)}$ is constant throughout the overlap although the two
constant values are not necessarily the same. In fact, the `gauge'
transformation (\ref{gaugetransformation}) supplies the link
between the values of these constants via
\begin{equation}\label{MvtoMu}
\hat{M}^{+}(b,t)=\ck^{-1}\partial_t\ck+\ck^{-1}\hat{M}^{-}(a,t)\ck\,
.
\end{equation}
Next, imagine shrinking the overlap region by allowing its
endpoints $a$, $b$ to approach the defect location at $x=0$. In
these circumstances, the  strategy used before, to obtain the
conserved charges in the bulk, can be applied. For example, the
analogue of (\ref{order1overl}) is
\begin{eqnarray}
\fl &&\hat{A}_x^{-}+\hat{A}_x^{+}
=-\frac{1}{\lambda}\left[(\bar{u}u)_t
+(\bar{v}v)_t+\rmi(u\hat{B}_x^{-}+\bar{u}
\hat{C}_x^{-})_x+\rmi(u\hat{B}_x^{+}+\bar{u}
\hat{C}_x^{+})_x\right.\nn\\
\fl &&\left. \ \ \ \ \ \ \ \ \ \ \ \ \ \ \ \ \ \ \ \ \ -\rmi(u_x
\hat{B}_x^{-}+\bar{u}_x \hat{C}_x^{-})-\rmi(u_x
\hat{B}_x^{+}+\bar{u}_x \hat{C}_x^{+})\right],
\end{eqnarray}
Integrating over  $-\infty<x <\infty$ gives
\begin{eqnarray}\label{order1overlwithdefect}
\fl
&&\left[\hat{A}^{-}-(\ck^{-1}\partial_t\ck+\ck^{-1}\hat{M}^{-}\ck)_{11}\right]_{x=0}=-\frac{1}{\lambda}
\left[\frac{\partial}{\partial t}\int^{0}_{-\infty}\rmd
x\,(\bar{u}u)+\frac{\partial}{\partial
t}\int^{\infty}_{0}\rmd x\,(\bar{v}v)\right]\nn\\
\fl &&\  -\frac{1}{\lambda} \left[\rmi(u\hat{B}_x^{-}+\bar{u}
\hat{C}_x^{-})-\rmi
u(\ck^{-1}\partial_t\ck+\ck^{-1}\hat{M}^{-}\ck)_{12}-\rmi
\bar{u}(\ck^{-1}\partial_t\ck+\ck^{-1}\hat{M}^{-}\ck)_{21}\right]_{x=0}\nn\\
\fl &&\ \ \ \ \ \ \ \ \ \ \ \ \ \ +\frac{1}{\lambda}
\left[\int^{0}_{-\infty}\rmd x\,\rmi(u_x \hat{B}_x^{-}+\bar{u}_x
\hat{C}_x^{-})+\int^{\infty}_{0}\rmd x\,\rmi(u_x
\hat{B}_x^{+}+\bar{u}_x \hat{C}_x^{+})\right]\, ,
\end{eqnarray}
where the elements of the matrix (\ref{MvtoMu}) appear. In turn,
these can be obtained using (\ref{modLpNLSu}) and (\ref{NLSk}).
Doing so, the first two lines of the integrated expansion
(\ref{order1overlwithdefect}) can be calculated and,
schematically, the result is the following
\begin{eqnarray}\label{order1overlwithdefect2}
\fl
&&\frac{1}{\lambda^0}\left[\cdots\right]_{x=0}+\frac{1}{\lambda}
\left\{\frac{\partial}{\partial t}\left(\int^{0}_{-\infty}\rmd
x\,(\bar{u}u)+\int^{\infty}_{0}\rmd x\,(\bar{v}v)\right)+\left[
-\frac{\partial\Omega}{\partial t}
+\cdots\right]_{x=0}\right\}+\frac{1}{\lambda^2}\left[\cdots\right]_{x=0}\nn\\
\fl &&\ \ \ \ \ \ \ \ \ \ \ \ \ \ \ \ -\frac{1}{\lambda}
\left[\int^{0}_{-\infty}\rmd x\,\rmi(u_x \hat{B}_x^{-}+\bar{u}_x
\hat{C}_x^{-})+\int^{\infty}_{0}\rmd x\,\rmi(u_x
\hat{B}_x^{+}+\bar{u}_x \hat{C}_x^{+})\right]=0\, ,
\end{eqnarray}
where the unwritten parts contain terms in $u$, $v$ and their
spatial derivatives. These terms cancel out after a lengthy calculation.
 In the first line of
(\ref{order1overlwithdefect2}) it is possible to recognize the
correction due to the defect, which has to be introduced to make
the `number' charge conserved (\ref{probderv}). The iteration
process has been checked up to  order $1/\lambda^2$ confirming the
correction to the momentum already indicated in (\ref{Qdot}). The
details of this verification are straightforward and will be
omitted.

It should be clear that these arguments can be repeated for an
arbitrary number of jump-defects placed along the $x$-axis. Each
defect is treated locally, introduces its own parameter, and
further complicates the generating functional (\ref{Qgenerator})
for the conserved quantities.

\section{A single soliton meeting a defect}

It is interesting to investigate what happens as a soliton
approaches a jump-defect. With the above normalisation for the
nonlinear term in the field equation, a one-soliton solution is
given by
\begin{equation}\label{NLSsoliton}
 \fl   u= \frac{2a\, EF}{1+E^2}\,, \qquad E=\exp{\left[a(x-2ct)\right]}\,, \qquad F=
    \exp{\left[\rmi\left(cx +\left(a^2 -c^2\right)t\right)\right]}
\end{equation}
where $a>0$ and $c$ are free, real parameters. For general values
of these parameters, the modulus and phase of the soliton travel
with different speeds. A slightly more general solution is
obtained by shifting the origin of $x$, for example, and
multiplying $u$ by an arbitrary phase.


When there is a defect,  the field $v$ on the other side of it is
taken to be given by a similar expression but with different
shifts in the modulus and the phase components; in other words,
take $v$ to be
\begin{equation}\label{}
    v= \frac{2a\, pq\, EF}{1+p^2\, E^2}\, .
\end{equation}
In view of the form of the defect conditions it is also reasonable
to suppose $p$ is real. Then, the first miracle which  should occur
concerns the square root in the defect conditions whose argument
will have to be a perfect square. For this one already requires the following
two relations between $p$ and $q$:
\begin{equation}\label{}
    \bar q q =1,\qquad \frac{|1-pq|^4}{(1-p^2)^2}=\frac{\alpha^2}{a^2}\, .
\end{equation}
Then, checking the second of equations (\ref{NLSBacklund})
requires
\begin{equation}\label{}
    pq=\frac{a +{\rmi c}-\alpha}{a + \rmi c+\alpha}\, ,
\end{equation}
and therefore
\begin{equation}\label{pandq}
    p^2=\frac{(a-\alpha)^2 +c^2}{(a+\alpha)^2 +c^2}, \qquad q= \frac{a-\alpha+\rmi c}
    {|a-\alpha+\rmi c|}\, \frac{a+\alpha-\rmi c}{|a+\alpha+\rmi c|}\, .
\end{equation}
The latter is quite a nice expression for $p$ since it can never
vanish (for $c\ne 0$), and it approaches unity as $c\rightarrow\infty$. Clearly, the
soliton cannot be `eaten' by the defect and the faster it travels the less it is
affected. Also, $p$ is not sensitive to the sign of
$c$, but changing the sign of $c$ replaces $q$ by $\bar q$. As
$|\alpha|\rightarrow\infty$ the parameter $q\rightarrow -1$,
indicating that when the defect parameter is sufficiently large the
soliton will have inverted its shape when it emerges from the
defect (but, since $p\rightarrow 1$, the inverted soliton will not
be significantly delayed). As $\alpha\rightarrow 0$, both
$p,q\rightarrow 1$ and the effect of the discontinuity disappears,
as it should.
Notice, the above collection of formulae refer to the positive
square root $\Omega$; changing the sign of the square root will
require changing $\alpha \rightarrow -\alpha$ in the above
expressions.

Overall, the picture for NLS is quite similar to that discovered
previously for sine-Gordon except that in the sine-Gordon case a
soliton may be absorbed by the defect, or converted to an
anti-soliton \cite{bczlandau} according to the choice of defect
parameter. These possibilities cannot occur for NLS.

\section{The two-soliton solution and a defect}

One would expect that a pair of initially widely separated
solitons approaching the defect should pass through it
independently, each experiencing the jump-defect as if the other
soliton were not there. Indeed this is the case although to verify
it explicitly is a formidable calculation requiring the use of
Mathematica or Maple. In this section the principal steps in this
verification will be outlined.

The multi-soliton solution has been given in closed form by
Zakharov and Shabat \cite{Zak72} (see also \cite{FadTak}) in terms
of the scattering data of the linear problem associated with the
NLS. This data is a set of complex constants
${\lambda_j=c_j+\rmi\,a_j\;\;a_j>0}$ and functions
${\Gamma_j(x,t)=\gamma_j\,\exp{i(\lambda_j x-\lambda_j^2 t)}}$
where $\gamma_j$ provides the initial position and phase of the
soliton labelled $j$. The two-soliton solution is then described
as follows

\begin{equation}\label{twosolitonsolution}
\fl \frac{u}{2}=\frac{\,a_1\Lambda_1\Gamma_1+
\,a_2\Lambda_2\Gamma_2+
\,a_1\overline{\Lambda}_1\,\Gamma_1\,|\Gamma_2|^2
+\,a_2\overline{\Lambda}_2\,\Gamma_2\,|\Gamma_1|^2 }
{1+|\Gamma_1|^2|\Gamma_2|^2+|\Lambda_1|^2|\Gamma_1|^2
+|\Lambda_2|^2|\Gamma_2|^2-4\,a_1 a_2
(\Gamma_1\overline{\Gamma}_2+\overline{\Gamma}_1\Gamma_2)/
|\lambda_1-\lambda_2|^2}
\end{equation}
where
\begin{equation}
\Lambda_2=\frac{(\lambda_2-\overline{\lambda}_1)}{(\lambda_2-\lambda_1)},\qquad
\Lambda_1=\frac{(\lambda_1-\overline{\lambda}_2)}{(\lambda_1-\lambda_2)}.
\end{equation}
Setting ${c=(c_1-c_2)}$, a solution of this type on the left of
the defect can be written conveniently as:
\begin{equation}\label{twosolitonftfullsolutionu}
\fl \frac{u}{2}=\frac{a_1E_1 F_1[\Delta_+ (1+E_2^2)-2\rmi c a_2
(1-E_2^2)]-a_2E_2F_2[\Delta_-(1+E_1^2)-2\rmi c
a_1(1-E_1^2)]}{(c^2+\delta_-^2)(1+E_1^2
E_2^2)+(c^2+\delta_+^2)(E_1^2+E_2^2)-4\,a_1 a_2\,E_1
E_2(F_1^2+F_2^2)/F_1F_2}
\end{equation}
where $\delta_+=(a_1+a_2)$, $\delta_-=(a_1-a_2)$,
$\Delta_{\pm}=\delta_+\delta_-\pm c^2$ and
\begin{equation}\label{notationforEF}
\fl E_j=\exp{[a_j(x-c_jt)-a_j\,x_{0j}]},\qquad
F_j=\exp{[\rmi(c_j\,x+(a_j^2-c_j^2)t+\phi_{0j})]} \quad j=1,2.
\end{equation}
Note that expression (\ref{twosolitonftfullsolutionu}) contains
the one-soliton solution (\ref{NLSsoliton}) on setting, for
instance, $a_2=c_2=0$.

To the right of the defect the expression
(\ref{twosolitonftfullsolutionu}) will be modified as follows
\begin{equation}\label{twosolitonftfullsolutionv}
\fl \frac{v}{2}=\frac{a_1G_1[\Delta_+ (1+H_2^2)-2\rmi c a_2
(1-H_2^2)]-a_2G_2[\Delta_-(1+H_1^2)-2\rmi c
a_1(1-H_1^2)]}{(c^2+\delta_-^2)(1+H_1^2
H_2^2)+(c^2+\delta_+^2)(H_1^2+H_2^2)-4\,a_1
a_2(G_1^2H_2^2+G_2^2H_1^2)/(G_1G_2)}
\end{equation}
where $G_j=z_jE_j F_j$, $H^2_j=w_jE^2_j$ and $w_j=p_j^2$,
$z_j=p_j\,q_j$ with $j=1,2$ and $|q_1|=|q_2|=1$. The constants
$p_j,\,\,q_j$ represent the delays in position and phase,
respectively, for the two solitons residing inside the solution
$v$, by analogy with the notation used for the one-soliton
solution on each side of the defect.

The first step in exploring the consequences of the defect
conditions is to find the circumstances under which the argument
of the square root $\Omega=(\alpha^2-|u-v|^2)^{1/2}$ is a perfect
square. The most general polynomial whose square can match the
argument of the square root must have the following form,
\begin{eqnarray}\label{polynomial}
\fl \Omega&=&(a+b_1\,E_1^2+b_2\,E_2^2+d_1\,E_1^4+d_2\,E_2^4+e_1\,E_1^4E_2^2+e_2\,E_1^2E_2^4\nn\\
\fl &&\ \ \ \ \ \ +\,g_1
\,E_1^3E_2+g_2\,E_1E_2^3+h\,E_1E_2+r\,E_1^3E_2^3+s\,E_1^4E_2^4+f\,E_1^2
E_2^2).
\end{eqnarray}
The coefficients of $\Omega^2$ are vastly overdetermined and it is
expected that just a few relationships among the parameters will
suffice. Indeed, this turns out to be the case, and the following
constraints on the constants $w_j,\,z_j$ are all that is required,
\begin{equation}\label{constraint1}
[a_j(z_j-1)(w_j-z_j)]^2=[\alpha\,(1-w_j)\, z_j]^2 \qquad j=1,2
\end{equation}
\begin{eqnarray}\label{constraint2}
\fl (a_1+a_2-\rmi c)(w_1-z_1)(z_2-1)&=&2\,\alpha \,(z_1-w_1z_2)\qquad
a_i\neq 0,\quad i=1,2\nonumber\\
\fl \ \ (a_1-a_2+\rmi c)(z_1-1)(z_2-1)&=&2\,\alpha\,(z_1-z_2)\qquad
\ \ \ a_i\neq 0,\quad i=1,2\,.
\end{eqnarray}
Note that, contrary to relations (\ref{constraint1}), the
expressions (\ref{constraint2}) are not real and therefore their
complex conjugate partners have also to be taken into account
\begin{eqnarray}\label{constraint2cc}
 \fl \ \ (a_1+a_2+\rmi c)(z_1-1)(w_2-z_2)&=&2\,\alpha
\,(z_2-w_2z_1)\qquad
\ \ \  a_i\neq 0,\quad i=1,2\nonumber\\
\fl (a_1-a_2-\rmi c)(w_1-z_1)(w_2-z_2)&=&2\,\alpha
\,(w_1z_2-w_2z_1)\qquad a_i\neq 0,\quad i=1,2\,.
\end{eqnarray}
Implementing these
constraints, the second part of the defect condition
 (\ref{NLSBacklund}) is satisfied provided the
following two relations hold
\begin{equation}
z_j=\frac{a_j-\alpha+\rmi\,c_j}{a_j+\alpha+\rmi\,c_j} \qquad j=1,2
\end{equation}
and therefore
\begin{equation}
w_j=\frac{(a_j-\alpha)^2+c_j^2}{(a_j+\alpha)^2+c_j^2} \qquad
j=1,2\,.
\end{equation}
Matching the conditions in this way demonstrates that solitons are
transmitted through the defect independently of one another. It is
expected this property should hold for arbitrary numbers of
solitons although there is no general proof of that yet.

In the sine-Gordon model it has been remarked that the delay
experienced by a soliton passing a defect is actually the square
root of the delay that would be experienced by the same  soliton
passing another whose rapidity was equal to the defect parameter
\cite{Bow05}. Using the following change of variables suggested in
\cite{Zak72} (see also \cite{FadTak}) when ${c_1>c_2}$,
\begin{equation}\label{changeofvariablesft}
\Gamma_1^{+}=\Gamma_1 \Lambda_1\qquad
\Gamma_1^{-}=\frac{\Gamma_1}{\Lambda_1}\qquad
\Gamma_2^{+}=\frac{\Gamma_2}{\Lambda_2}\qquad
\Gamma_2^{-}=\Gamma_2 \Lambda_2\, ,
\end{equation}
the two soliton solution (\ref{twosolitonsolution}) may be
rewritten as follows
\begin{equation}\label{twosolitonsolutionft}
\fl  \frac{u}{2}=\frac{\,a_1\Gamma_1^{+}+\,a_2\Gamma_2^{-}+
\,a_1\Gamma_1^{-}|\Gamma_2^{-}|^2+\,a_2\Gamma_2^{+}|\Gamma_1^{+}|^2}
{1+|\Gamma_1^{+}|^2+
|\Gamma_2^{-}|^2+|\Gamma_1^{+}|^2|\Gamma_2^{+}|^2-8\,a_1 a_2{\rm
Re}[(\Gamma_1^+\bar{\Gamma}_2^{-})/(\lambda_1-\bar{\lambda}_2)^2]}.
\end{equation}
Note that the change of variables (\ref{changeofvariablesft})
always holds provided $\lambda_1\neq \lambda_2,\,\bar{\lambda}_2$.
Moreover \begin{equation}|\Gamma_1^{+}|^2|\Gamma_2^{+}|^2=
|\Gamma_1^{-}|^2|\Gamma_2^{-}|^2=|\Gamma_1|^2|\Gamma_2|^2.\end{equation}

Functions $\Gamma_j^{\pm}(x,t)$ are defined as follows
\begin{equation}
\Gamma_j^{\pm}(x,t)=\gamma_j^{\pm}\;\exp{[\rmi(\lambda_j
x-\lambda_j^2 t)]},\qquad
\gamma_j^{\pm}=\exp{(a_jx_{0j}^{\pm}+\rmi\phi_{0j}^{\pm})}
\end{equation}
where now
\begin{eqnarray}
&&\Delta p_0=\exp{[-a_1(x_{01}^{+}-x_{01}^{-})]}
=\exp{[a_2(x_{02}^{+}-x_{02}^{-})]} ,\nonumber\\
&&\Delta q_{01}=\exp{[-\rmi(\phi_{01}^{+}-\phi_{01}^{-})]} ,\qquad
\Delta q_{02}=\exp{[\rmi(\phi_{02}^{+}-\phi_{02}^{-})]}
\end{eqnarray}
are the two-soliton scattering data. In fact, looking at the limit
$t=\pm \infty$ of the two-soliton solution
(\ref{twosolitonsolutionft}), for instance along the first soliton
trajectory $x-c_1 t={\rm constant}$, it is possible to obtain the
following one-soliton solutions ($c_1>c_2$)
\begin{equation}
u|_{\,t\rightarrow
\infty}=\frac{2\,a_1\,\Gamma_1^{+}}{1+|\Gamma_1^{+}|^2}, \qquad
u|_{\,t\rightarrow
-\infty}=\frac{2\,a_1\,\Gamma_1^{-}}{1+|\Gamma_1^{-}|^2}.
\end{equation}
Using (\ref{changeofvariablesft}), explicit expressions for the
scattering data $\Delta p_0$, $\Delta q_{01}$ and $\Delta q_{02}$
are:
\begin{equation}
\Delta p_0=\frac{(a_1-a_2)^2+c^2}{(a_1+a_2)^2+c^2},
\end{equation}
\begin{eqnarray}
(\Delta q_{01})^{-1}&=&\exp{[\rmi(\phi_{01}^{+}-\phi_{01}^{-})]}=
\frac{a_1^2-a_2^2+2\rmi a_2 c+c^2}{a_1^2-a_2^2-2\rmi
a_2c+c^2},\nn\\
\Delta q_{02}&=&\frac{a_2^2-a_1^2+2\rmi a_1
c+c^2}{a_2^2-a_1^2-2\rmi a_1c+c^2}.
\end{eqnarray}
Note that $(\Delta q_{01})^{-1}$ can be also written as follows
\begin{equation}
(\Delta q_{01})^{-1}=
\frac{(a_1^2-a_2^2)^2+c^4+2\,c^2a_1^2-6\,c^2a_2^2+4\,\rmi
a_2(c^2+a_1^2-a_2^2)} {(a_1^2-a_2^2)^2+c^4+2\,c^2(a_1^2+a_2^2)}.
\end{equation}
It is then easy to check that $\Delta p_0$ and $(\Delta
q_{01})^{-1}$ coincide with the squares of the shifts in the
modulus $p$ and  the phase $q$ (\ref{pandq}) experienced by the
soliton progressing through a defect, provided $c_2=0$ and
$a_2=\alpha$, with $\alpha$ being the defect parameter.

\section{Bound states}

With a delta function impurity, in a free field situation with a
suitable coupling, there will be a bound state. Curiously, this is
not the case for the free field limit of the sine-Gordon model
with the jump-defect conditions provided by (\ref{sGdefect}), nor
is it the case for the full sine-Gordon model with a jump-defect.
In the sine-Gordon model this type of purely transmitting defect
does not permit bound states for any value of the coupling
$\sigma$. Allowing a jump-defect in the quantum sine-Gordon model
does have interesting effects, however, and it appears a defect
can be excited, albeit with a finite decay width. This is the
quantum analogue of the classical property permitting a
jump-defect to swallow a soliton \cite{Bow05}.

On the other hand, a jump-defect in the NLS model certainly has
bound states associated with it. The first indication of this
arises using the linear, potential-free, Schr\"odinger equations
and the linearised versions of the defect conditions
(\ref{NLSBacklund})
\begin{eqnarray}\label{NLSlinear}
\nn u_x&=&-\frac{\rmi}{2\alpha}(u_t-v_t) +\frac{\alpha}{2}(u+v)\\
v_x&=&-\frac{\rmi}{2\alpha}(u_t-v_t) -\frac{\alpha}{2}(u+v).
\end{eqnarray}
It is easy to check these allow the travelling wave solution
\begin{equation}\label{travellingwave}
  \fl  u=u_0\, \exp{(-\rmi k^2 t +\rmi kx)}, \qquad v=v_0\, \exp{(-\rmi k^2 t +\rmi kx)} ,\qquad v_0=
    \frac{k+\rmi\alpha}{k-\rmi\alpha}\, u_0,
\end{equation}
where $k$ is real, from which the existence of bound states
associated with the defect may be deduced. If $\alpha >0\ (<0)$,
(\ref{travellingwave}) indicates a bound state when $k=\rmi\alpha\
(-\rmi\alpha)$ for which there are the square integrable solutions
\begin{equation}\label{}
   \fl u=0,\quad v=v_0\,\exp{(\rmi\alpha^2 t -\alpha x)}\,;
    \qquad (u=u_0\,\exp{(\rmi\alpha^2 t +\alpha x)},\quad v=0),
\end{equation}
respectively.

Interestingly, these bound states have their counterparts in the
fully nonlinear version of the model. For example, taking $\alpha$
to be positive, with $a=\alpha,\ c=0$ in (\ref{NLSsoliton}), the
field configurations
\begin{equation}\label{NLSboundstate}
    u=\frac{2\alpha\, \rme^{\alpha (x-x_0)}\, \rme^{\rmi\alpha^2 t}}
    {1+\rme^{2\alpha (x-x_0)}},\qquad v=0,
\end{equation}
in the regions $x<0$ and $x>0$, respectively, satisfy the jump
defect conditions. If the parameter $x_0=0$, the quantity $\Omega$
actually vanishes at the defect, as does the momentum of course,
while ${N}=\alpha$ and $E=-\alpha^3/3$ are exactly half the values
the number and energy would have had for a static breather
solution (with $a=\alpha,\ c=0$) in the bulk. In fact, the latter
values for $N$ and $E$ remain correct if $x_0\ne 0$; in those
cases $\Omega$ does not vanish but compensates the bulk integrals.
Note, this solution is entirely consistent with the expressions
discovered before in (\ref{pandq}), noting that if $c=0$ then $p$
vanishes precisely when $a=\alpha$.

\section{The modified Korteweg-de Vries}

One might wonder if the modified KdV (mKdV) equation, or the KdV
itself, both of which certainly possesses B\"acklund
transformations, also allow discontinuous solutions, or defects,
provided suitable conditions are imposed.

In the bulk, a mKdV equation is \cite{D&J}
\begin{equation}\label{mKdV}
    v_t - 6 v^2v_x -v_{xxx}=0,
\end{equation}
or,  setting $v=p_x$,
\begin{equation}\label{pmKdV}
p_{xt}-6p_x^2p_{xx}-p_{xxxx}=0,
\end{equation}
the latter being suitable for a Lagrangian description with
Lagrangian density
\begin{equation}\label{lagrangianmKdV}
    {\cal L}=\frac{1}{2}p_xp_t-\frac{1}{2}(p_x)^4+\frac{1}{2}(p_{xx})^2.
\end{equation}
Integrating (\ref{pmKdV}) once with respect to $x$, and assuming
all derivatives are asymptotically vanishing gives an alternative:
\begin{equation}\label{alternativepmKdV}
    p_t-2p_x^3-p_{xxx}=0.
\end{equation}

There is another version of mKdV (obtained by the replacement
$v\rightarrow \rmi v$) in which the middle term on the left hand
side of (\ref{mKdV}) changes sign. However, it is equation
(\ref{mKdV}) which allows real soliton solutions (see below), not
the alternative version \cite{Ablowitz73}.


Over the whole line, the quantity
\begin{equation}\label{momentummKdV}
    P=\frac{1}{2}\int_{-\infty}^\infty \, \rmd x\,  p_x^2,
\end{equation}
is conserved (with the usual assumptions at $\pm \infty$) because
\begin{equation}\label{continuitymP}
    \left(\frac{p_x^2}{2}\right)_t= \left(p_t p_x-\frac{1}{2}p_x^4-\frac{p_{xx}^2}{2}\right)_x,
\end{equation}
where the last expression made use of the alternative equation
(\ref{alternativepmKdV}).

Next, suppose there is a `defect' at $x=0$, with fields $p,q$ on
either side of it. The quantity $P$ defined by
\begin{equation}\label{}
    P=\frac{1}{2}\int_{-\infty}^0 \, \rmd x\,  p_x^2+\frac{1}{2}\int_{0}^\infty \,
    \rmd x\,  q_x^2
\end{equation} will not be conserved but, as a consequence of (\ref{continuitymP}), its
time derivative will be related to a boundary term as follows,
\begin{equation}\label{mtderivativeP}
    P_t=\left(p_t p_x-\frac{p_x^4}{2}-\frac{p_{xx}^2}{2}\right)_{x=0} -
    \left(q_t q_x-\frac{q_x^4}{2}-\frac{q_{xx}^2}{2}\right)_{x=0}.
\end{equation}
The question is how to write the latter as a time derivative of
the fields and their derivatives evaluated at $x=0$.

To see where the defect conditions are coming from, it is
necessary to return and consider the complete action
\begin{equation}\label{action}
    {\cal A}= \int \rmd t \left\{ \int_{-\infty}^0 \rmd x\, {\cal L}(p) +{\bf B} +\int_0^\infty \rmd x\,
    {\cal L}(q)\right\},
\end{equation}
and its variation with respect to $p$ or $q$. Thus, for example, varying $p$ gives,
\begin{eqnarray}
 \fl  \delta A &=&  \int \rmd t \left\{ \int_{-\infty}^0 \rmd x\, \left(\frac{1}{2}\delta p_t p_x +
   \frac{1}{2}\delta p_x p_t -2 \delta p_xp_x^3 +\delta p_{xx} p_{xx}\right) \right.\\
\fl   &&\qquad \qquad \qquad \qquad \qquad + \left. \left(\delta
p_t
   \frac{\partial {\bf B}}{\partial p_t} +\delta p
   \frac{\partial {\bf B}}{\partial p} +\delta p_x
   \frac{\partial {\bf B}}{\partial p_x}\right)_{x=0}\right\},
\end{eqnarray}
which, on integrating the first term by parts with respect to $t$
and $x$, keeping the boundary terms in $x$, and setting the
variation to zero, leads to the field equation for $p$ together
with
\begin{equation}\label{}
 \fl   0=\left[ \delta p \left(\frac{1}{2} p_t - 2 p_x^3-p_{xxx} +\frac{\partial {\bf B}}{\partial p}
    -\frac{\partial}{\partial t}\frac{\partial {\bf B}}{\partial p_t}\right) +\delta p_x\left(
    p_{xx} + \frac{\partial {\bf B}}{\partial p_x}\right)\right]_{x=0}.
\end{equation}
At the defect, there is no necessity for  $\delta p$ and $\delta
p_x$ to be related but (\ref{alternativepmKdV}) can be used (in a
limiting sense) to eliminate $p_{xxx}$. Hence, the defect
conditions on $p$ should read
\begin{equation}\label{mpdefectconditions}
  0 = -\frac{1}{2} p_t +\frac{\partial {\bf B}}{\partial p}- \frac{\partial}{\partial t}
  \frac{\partial {\bf B}}{\partial p_t}\,, \qquad
 0 =  p_{xx} +\frac{\partial {\bf B}}{\partial p_x}\, .
\end{equation}
Similarly, the relations to be satisfied by $q$ at the defect are:
\begin{equation}\label{mqdefectconditions}
  0 = \frac{1}{2} q_t +\frac{\partial {\bf B}}{\partial q}- \frac{\partial}{\partial t}
  \frac{\partial {\bf B}}{\partial q_t}, \qquad
  0 = -q_{xx} +\frac{\partial {\bf B}}{\partial q_x}\, .
\end{equation}
The next step is to find a suitable defect term ${\bf B}$.
However, this is not quite straightforward and before doing so it
is worth making a remark.

It is tempting to assume the defect term depends only on $p,q$ and
their first derivatives (in fact this was tacitly assumed above).
However, there is no reason in principle why this term should not
depend on $p_{xx}$ and $q_{xx}$ since the field equations cannot
at the defect relate these derivatives to anything else (but note,
the same would not be true of $p_{xxx}$, or $q_{xxx}$). Moreover,
since the Lagrangian (\ref{lagrangianmKdV}) depends upon both
first and second derivatives, it should not be surprising that
defect conditions should also involve higher spatial derivatives.
If ${\bf B}$ does depend on these derivatives then there will be
two more defect conditions coming from varying the boundary term
alone, unbalanced by anything in the bulk; they are
\begin{equation}\label{newconditions}
    \frac{\partial\,{\bf B}}{\partial p_{xx}}=0,
    \qquad \frac{\partial\,{\bf B}}{\partial q_{xx}}=0.
\end{equation}


On the other hand, a B\"acklund transformation for the mKdV
equation was found long ago \cite{Lamb74, Wadati74} and can be
written in the following symmetrical manner:
\begin{eqnarray}\label{mKdVBacklund}
\fl \nn p_x+q_x&=&\alpha\sin(p-q),\\
\fl \nn p_t+q_t&=&\alpha\left(p_{xx}-q_{xx}\right)\cos(p-q) +
\alpha\left(p_x^2+q_x^2\right)\sin(p-q)\\
\fl \phantom{p_t+q_t\ }&
=&\alpha\left(p_{xx}-q_{xx}\right)\cos(p-q)-2\alpha
p_xq_x\sin(p-q)+ \alpha^3\sin^3(p-q).
\end{eqnarray}
The parameter $\alpha$ is arbitrary and the pair of equations is
symmetrical under interchanging $p$ and $q$, and simultaneously
making the replacement $\alpha\rightarrow -\alpha$. In the bulk,
the derivative of the first of the pair (\ref{mKdVBacklund}) is
\begin{equation}\label{extramKdV}
    p_{xx}+q_{xx}=\alpha\left(p_x-q_x\right)\cos(p-q),
\end{equation}
which will provide the extra equation coming from
(\ref{newconditions}) (see below). Note that, under the
circumstances being explored here, the $x$-derivatives are frozen,
therefore (\ref{extramKdV}) cannot be a consequence of the first
of conditions (\ref{mKdVBacklund}).

Using (\ref{mKdVBacklund}) and (\ref{extramKdV}) the expression
(\ref{mtderivativeP}) can be simplified to
\begin{equation}\label{}
    P_t=\frac{d}{dt}\left[-\alpha\cos(p-q)\right]_{x=0},
\end{equation}
and therefore, \begin{equation}P+\alpha[\cos(p-q)
-1]_{x=0}\end{equation} is conserved. The constant has been chosen
to ensure the momentum stored at the defect is zero when there is
no discontinuity. It is worth noting that because of the
discontinuity the charge that in the bulk is obtained from the
density ${\cal N}=p_x$ is not conserved. The same could be said
for the KdV model discussed in \sref{section10}.

As previously, it is useful to put
\begin{equation}{\bf B}=  \frac{1}{4}(qp_t-pq_t)-{\cal B}\end{equation}
and the defect conditions may be rewritten
\begin{equation}\label{mpdefectconditions}
\fl    \left(p_t +q_t\right)=-2\frac{\partial {\cal B}}{\partial p}=
   2\frac{\partial {\cal B}}{\partial q}\,; \quad
  p_{xx} =\frac{\partial {\cal B}}{\partial p_x},\quad q_{xx} =
 -\frac{\partial {\cal B}}{\partial q_x}\,;\quad 0=\frac{\partial {\cal B}}{\partial p_{xx}}
=
 \frac{\partial {\cal B}}{\partial q_{xx}}\,.
\end{equation}
Putting everything together, a suitable choice of ${\cal B}$ appears to be to take
\begin{eqnarray}\label{mKdVB}
\fl \nn    {\cal B}&=&\frac{1}{2}\left(p_{xx}-q_{xx}\right)\left(p_x+q_x-\alpha\sin(p-q)\right)\\
 \fl   && \ \ \ \ \ \ \ + \frac{\alpha}{6}\cos(p-q)
    \left[p_x^2+q_x^2-4p_xq_x+\alpha\left(p_x+q_x\right)\sin(p-q)
    +\alpha^2\right].
\end{eqnarray}
Using (\ref{mKdVB}) the defect conditions are exactly equivalent
to what would be the B\"acklund transformation in the bulk
together with (\ref{extramKdV}) were the conditions not frozen at
$x=0$. Finally, it is worth remarking that  the mKdV defect
potential is not a simple function of the two fields $u$ and $v$
and the parameter $\alpha$ but rather of the `potentials' $p$ and
$q$.

\section{The mKdV soliton passing through a defect}

The single soliton for the mKdV equation can be conveniently
written in terms of $p$ in the form
\begin{equation}\label{mKdVsoliton}
\rme^{\rmi p}=\frac{1+\rmi E}{1-\rmi E}\,, \qquad E=\exp{\left[a
(x-x_0+a^2 t)\right]},
\end{equation}
where $a$ and $\exp{(x_0)}$ are both real parameters. With the
conventions adopted in the last section, the soliton moves from
right to left along the $x$-axis. Since the derivative of the
above expression is either always positive ($E>0$) or always
negative ($E<0$), there is also a matching  anti-soliton obtained
by replacing $E\rightarrow -E$ in (\ref{mKdVsoliton}). This is
easily seen, on noting
\begin{equation}p_x=\frac{2a E}{1+E^2}\,.\end{equation}
If there is a defect, the soliton on the other side of it will
have the form
\begin{equation}\label{}
\rme^{\rmi q}=\frac{1+\rmi zE}{1-\rmi zE}\,, \qquad E=\exp{\left[a
(x-x_0+a^2 t)\right]},
\end{equation}
where $z$ is a parameter to be determined by the
defect condition.

The first of the defect conditions readily reveals that
\begin{equation}\label{zmKdV}
    z=\frac{\alpha -a}{\alpha+a}\,.
\end{equation}
This appears to suggest that a slow soliton is not affected much
whereas a fast soliton has $z<0$. This means that a fast soliton
flips to an anti-soliton ($E\rightarrow -E$ in the expression
(\ref{mKdVsoliton})). When $a=\alpha$, the soliton is eaten. This
behaviour is very similar to that of a soliton meeting a defect in
the sine-Gordon model where all these effects are similarly
apparent. Despite the non-locality, there appears to be nothing
particularly pathological about this case.  As a final remark in
this section, it is clear the effect of the defect disappears as
$\alpha\rightarrow \infty$.

\section{The Korteweg-de Vries equation}
\label{section10}

It is not difficult to repeat these steps for the KdV equation. However, there does
seem to be some curiously different behaviour that will become apparent as this
section proceeds.

In the bulk, the KdV equation is \cite{D&J}
\begin{equation}\label{KdV}
    u_t - 6 uu_x +u_{xxx}=0,
\end{equation}
or,  setting $u=p_x$,
\begin{equation}\label{pKdV}
p_{xt}-6p_xp_{xx}+p_{xxxx}=0,
\end{equation}
the latter being suitable for a Lagrangian description with
Lagrangian density
\begin{equation}\label{lagrangianKdV}
    {\cal L}=\frac{1}{2}p_xp_t-(p_x)^3-\frac{1}{2}(p_{xx})^2.
\end{equation}
Integrating (\ref{pKdV}) once with respect to $x$, and assuming
all derivatives are asymptotically vanishing gives an alternative:
\begin{equation}\label{alternativepKdV}
    p_t-3p_x^2+p_{xxx}=0.
\end{equation}
Over the whole line, the quantity
\begin{equation}\label{momentumKdV}
    P=\frac{1}{2}\int_{-\infty}^\infty \, \rmd x\,  (p_x)^2,
\end{equation}
is conserved (with the usual assumptions at $\pm \infty$) because
\begin{equation}\label{continuityP}
    \left(\frac{p_x^2}{2}\right)_t=\left(2p_x^3 - p_xp_{xxx} +
    \frac{p_{xx}^2}{2}\right)_x = \left(p_t p_x-p_x^3+\frac{p_{xx}^2}{2}\right)_x,
\end{equation}
where the last expression made use of the alternative equation
(\ref{alternativepKdV}).

Next, suppose there is a `defect' at $x=0$, with fields $p,q$ on
either side of it. The quantity $P$ defined by
\begin{equation}\label{}
    P=\frac{1}{2}\int_{-\infty}^0 \, \rmd x\,  p_x^2+\frac{1}{2}\int_{0}^\infty \,
    \rmd x\,  q_x^2
\end{equation} is not conserved but, as a consequence of (\ref{continuityP}), its
time derivative will be related to a boundary term as follows,
\begin{equation}\label{tderivativeP}
    P_t=\left(p_t p_x-p_x^3+\frac{p_{xx}^2}{2}\right)_{x=0} -
    \left(q_t q_x-q_x^3+\frac{q_{xx}^2}{2}\right)_{x=0}.
\end{equation}
The question is how to write the latter as a time derivative of
the fields or their derivatives evaluated at $x=0$.

The right hand side of (\ref{tderivativeP}) can be organised a
little differently (dropping the explicit reference to the point
$x=0$, which is to be understood from now on) to,
\begin{equation}\label{}
 \fl   -(p_x-q_x)(p_x^2 +p_xq_x +q_x^2)+\frac{1}{2}(p_{xx}-q_{xx})(p_{xx}+q_{xx})+p_tp_x-q_tq_x,
\end{equation}
and then simplified by setting\begin{eqnarray}
 \nonumber\label{conditionone}\ \ \ \ \ \ \ p_{xx}+q_{xx}& = & (p-q)(p_x-q_x)  \\
   p_x^2+p_x q_x +q_x^2&=&\frac{1}{2}\left[p_t+q_t +(p-q)(p_{xx}-q_{xx})\right],
\end{eqnarray}
to get
\begin{equation}\label{}
    \frac{1}{2}(p_t-q_t)(p_x+q_x).
\end{equation}
Finally, setting in addition
\begin{equation}\label{conditiontwo}
    p_x+q_x=2\alpha +\frac{1}{2}(p-q)^2,
\end{equation}
one finds
\begin{equation}\label{compensatingterm}
P_t=\frac{d}{dt}\left[\alpha(p-q)+\frac{1}{12}(p-q)^3\right]_{x=0},
\end{equation}
where the parameter $\alpha$ is arbitrary. The first of the
conditions (\ref{conditionone}) would follow from
(\ref{conditiontwo}) in the bulk but, here, as already mentioned
before, the $x$-derivatives are frozen. The second of equations
(\ref{conditionone}) together with (\ref{conditiontwo}) provides a
B\"acklund transformation for KdV in the bulk; at least in the
sense that if $p$ satisfies (\ref{pKdV}) so does $q$ (or
vice-versa), for any choice of $\alpha$. This is the form of
B\"acklund transformation constructed by Wahlquist and Estabrook \cite{Wahlquist}
for generating multi-soliton solutions to the KdV equation.

This time using the Lagrangian together with a defect contribution
(assuming the latter depends on $p$ and $q$ and their derivatives
$p_x,\ q_x$ and $p_{xx},\ q_{xx}$), leads naturally to the defect
conditions:
\begin{equation}\label{pdefectconditions}
  0 = -\frac{1}{2} p_t +\frac{\partial {\bf B}}{\partial p}- \frac{\partial}{\partial t}
  \frac{\partial {\bf B}}{\partial p_t}, \qquad
 0 = -p_{xx} +\frac{\partial {\bf B}}{\partial p_x}, \qquad 0=\frac{\partial {\bf B}}{\partial p_{xx}}\,.
\end{equation}
Similarly, the relations to be satisfied by $q$ at the defect are:
\begin{equation}\label{qdefectconditions}
  0 = \frac{1}{2} q_t +\frac{\partial {\bf B}}{\partial q}- \frac{\partial}{\partial t}
  \frac{\partial {\bf B}}{\partial q_t}, \qquad
 0 = q_{xx} +\frac{\partial {\bf B}}{\partial q_x}, \qquad
  0=\frac{\partial {\bf B}}{\partial q_{xx}}\,.
\end{equation}
A suitable choice for the jump-defect potential might be the
following
\begin{eqnarray}\label{newB}
\fl  \nn &&  {\bf B}=\frac{1}{4}\left(qp_t-pq_t\right)+ \left(p_x^2+p_x q_x +q_x^2\right)(p-q)
    +\frac{1}{2}\left(p_{xx}-q_{xx}\right)\left[p_x+q_x - 2\alpha -\frac{1}{2}(p-q)^2\right]\\
\fl   && \  -3\left(p_x+q_x\right)(p-q)\left(\alpha + \frac{1}{4}(p-q)^2\right)
+ 6\alpha^2(p-q)+2\alpha (p-q)^3+\frac{9}{40}(p-q)^5.
\end{eqnarray}
This looks quite complicated but it does not seem easy to find
anything simpler that would be able to provide the relations
needed. In this respect, both the last equations in
(\ref{pdefectconditions}), (\ref{qdefectconditions}) give
precisely the equation (\ref{conditiontwo}), while two of the
other conditions, still from (\ref{pdefectconditions}),
(\ref{qdefectconditions}), give
\begin{eqnarray}
 \fl p_{xx}&=&\phantom{-}\frac{1}{2}\left(p_{xx}-q_{xx}\right) +
  \left(2p_x+q_x\right)(p-q) -3\alpha (p-q)-\frac{3}{4}(p-q)^3, \\
\fl  q_{xx}&=&-\frac{1}{2}\left(p_{xx}-q_{xx}\right) -
\left(p_x+2q_x\right)(p-q)
  +3\alpha (p-q)+\frac{3}{4}(p-q)^3.
\end{eqnarray}
Adding these gives
\begin{equation}\label{}
    p_{xx}+q_{xx}=\left(p_x-q_x\right)(p-q),
\end{equation}
which, in the bulk, is just the derivative of
(\ref{conditiontwo}); subtracting them gives (\ref{conditiontwo})
again (the second derivatives exactly cancelling out).

Finally, the other pair of equations from
(\ref{pdefectconditions}), (\ref{qdefectconditions}) give
effectively the same condition, namely
\begin{equation}\label{}
    \frac{1}{2}\left(p_t+q_t\right)=-\frac{1}{2}\left(p_{xx}-q_{xx}\right)(p-q) +
    \left(p_x^2+p_xq_x+q_x^2\right).
\end{equation}
This is not quite straightforward to see and makes use of
(\ref{conditiontwo}) again. Indeed, the coefficients of the
polynomial part of ${\bf B}$ were chosen to ensure this worked
out.

With this particular choice of ${\bf B}$, the `momentum' is
compensated in precisely the manner envisaged before (in
(\ref{compensatingterm})), and
\begin{equation}\label{conservedP}
    P-\left[\alpha(p-q) +\frac{1}{12}(p-q)^3\right]_{x=0}
\end{equation}
is conserved. Moreover, the parameter $\alpha$ is completely free.

The next bulk conserved quantity, the `energy', has a density
\begin{equation}\label{}
  {\cal E}=\left[(p_x)^3+\frac{1}{2}(p_{xx})^2\right],
\end{equation}
It is not conserved but the defect contributes exactly
\begin{equation}\label{}
    E_t=\left(-\frac{d{\cal B}}{dt}\right)_{x=0},
\end{equation}
provided the defect potential (\ref{newB}) is written
\begin{equation}\label{defectpotentialprototype}
    {\bf B}=\frac{1}{4}(qp_t-pq_t)-{\cal B}.
\end{equation}
Hence, using the boundary conditions
\begin{equation}\label{conservedE}
 \fl   E+{\cal B}=E-\left[\left(p_x^2+p_x q_x +q_x^2\right)(p-q)+ (p-q)^3\left(\alpha+\frac{3}{20}
    (p-q)^2\right)\right]_{x=0}
\end{equation}
is conserved.

The situation is fairly similar to what was found for mKdV. It is
curious that neither of the compensating terms can be rewritten in
terms of the the original KdV variables $u$ on the left, or $v$ on
the right ($u=p_x,\ v=q_x$). The expressions for energy and momentum reveal that there is
nothing particularly special about a vanishing parameter $\alpha$.

In general, provided the defect potential ${\bf B}$ is chosen as
it is in (\ref{defectpotentialprototype}), the momentum is
conserved provided

\begin{equation}
\left(\frac{\partial {\cal B}}{\partial
p_x}\right)^2-\left(\frac{\partial {\cal B}}{\partial
q_x}\right)^2+(p_x-q_x)\left[2\frac{\partial{\cal B}}{\partial
p}+(p_x^2+q_x^2)(p_x+q_x)\right]=0,
\end{equation}
for mKdV and
\begin{equation}\label{momentumconsmKdV1}
\left(\frac{\partial {\cal B}}{\partial
p_x}\right)^2-\left(\frac{\partial {\cal B}}{\partial
q_x}\right)^2-(p_x-q_x)\left[2\frac{\partial{\cal B}}{\partial
p}+2(p_x^2+p_xq_x+q_x^2)\right]=0,
\end{equation}
for KdV, respectively, where
\begin{equation}
 \left(\frac{\partial {\cal B}}{\partial
p_x}\right)^2-\left(\frac{\partial {\cal B}}{\partial
q_x}\right)^2=(p_{xx}+q_{xx})(p_{xx}-q_{xx}),\quad
2\frac{\partial{\cal B}}{\partial p}=-(p_t+q_t),
\end{equation}
together with
\begin{equation}\label{momentumcons1}
(p_t-q_t)(p_x+q_x)=2\frac{d{\cal F}}{d t},
\end{equation}
for either model, where ${\cal F}$ is a function to be determined.

\section{KdV solitons and the defect}

A single soliton can be described conveniently by choosing
\begin{equation}\label{solitonp}
 \fl   p=p_0- \frac{2\,a E}{1+E}, \qquad E=\exp{\left[a(x-x_0-a^2 t)\right]}\equiv
    \rho\,
    \exp{\left[a(x-a^2 t)\right]},
\end{equation}
where $p_0,a,x_0$ are constants. In terms of the original variable
$u$ the solution has the well-known characteristic `bell-shaped' form
\begin{equation}\label{}
    u=p_x=-\frac{2\,a^2 E}{(1+E)^2}.
\end{equation}The parameter $a$ is real and the soliton is
independent of the sign of $a$ (since $p_x$ is the same whatever
the sign). Inevitably, the soliton moves to the right.

Then, it is not difficult to check that picking a similar form for
$q$
\begin{equation}\label{solitonq}
    q=q_0- \frac{2\,a\, \sigma E}{1+\sigma E}, \qquad E=\exp{\left[a(x-a^2 t)\right]},
\end{equation}
and redefining the defect parameter $\alpha=-\beta^2/4$, equation
(\ref{conditiontwo}) implies
\begin{equation}\label{}
    \left(p_0-q_0\right)^2=\beta^2, \qquad \sigma =\left(\frac{|\beta| -a}{|\beta|+a}
    \right)\, \rho.
\end{equation}
At least that is the case assuming the positive square root is
taken for $p_0-q_0$. The other equation (the second of
(\ref{conditionone})) is just an identity using only the fact
\begin{equation}\label{pminusq}
    p_0-q_0=a \,\left( \frac{\rho+\sigma}{\rho-\sigma}\right),
\end{equation}
as can be checked easily using an algebraic computing package.
Finally, the first of conditions (\ref{conditionone}) is also an
identity and can be checked directly. The relation (\ref{pminusq})
does not depend on taking square roots.

This implies a soliton encountering the defect, provided it is not
travelling too quickly, will be delayed. When $a$ approaches
$\beta$, $\sigma$ tends to zero and the soliton will
have been `eaten' (since $q=q_0$). There is a mystery if the
soliton is moving too quickly because the solution on the right of
the defect develops a singularity. This is puzzling for another
reason. If the defect parameter is taken to be very small then
only a very slow soliton will remain non-singular, meaning that in
the limit as the parameter goes to zero one cannot recover a
single bulk soliton because the speed of the non-singular soliton
would also have to approach zero. Similarly, in the limit of large
defect parameter $(p_0-q_0)^2\rightarrow\infty$, implying that the
effect of the defect on the soliton does not entirely disappear,
although the fields to either side of it will be the same (because
$p_x\rightarrow q_x$). In these respects, the KdV defect behaves
curiously
 differently from those encountered elsewhere.

One might wonder whether having  a single soliton on the left of
the defect and a double soliton on the right might provide some
hints towards solving the mystery. To explore this, consider the
following single one-solution on the left and two-soliton solution
on the right of the defect, respectively
\begin{eqnarray}\label{solitonpz}
 p&=&p_0- \frac{2\,a_2 \rho_2E_2}{1+\rho_2E_2},\nn \\
q&=&q_0+\frac{(a_1^2-a_2^2)(1+\rho_1E_1+\sigma\rho_2 E_2+\sigma\rho_1\rho_2 E_1E_2)}
    {(a_1-a_2)(1-\sigma\rho_1\rho_2 E_1E_2)-(a_1+a_2)(\rho_1 E_1-\sigma\rho_2 E_2)},
\end{eqnarray}
with
\begin{equation}
\rho_jE_j=\exp{\left[a_j(x-x_{0j}-a_j^2 t)\right]}\qquad j=1,2.
\end{equation}

As pointed out in \cite{Wahlquist}, a two-soliton solution is
non-singular provided one of the two component solitons
 is actually singular. For instance, in
(\ref{solitonpz}) suppose $a_2>a_1$, then for $q$ to be regular
$\rho_1>0$ and $\sigma\rho_2<0$ and consequently the faster
soliton is singular. Checking the B\"acklund equation
(\ref{conditiontwo}) for the solution (\ref{solitonpz}) reveals
the following
\begin{equation}
a_1^2=\beta^2,\qquad (p_0-q_0)^2=\beta^2, \qquad\sigma=1;
\end{equation}
therefore, provided the incoming soliton is regular ($\rho_2>0$),
the resulting two-soliton solution is singular. One could think
that the `fast' soliton is trapped by the defect since it does not
have enough energy to escape. On the other hand, if the
one-soliton solution on the left is singular ($\rho_2>0$) then the
resulting two-soliton solution would be regular (still with
$a_2>a_1$). The singularity of the incoming soliton can be kept to
the right of the defect by a suitable choice of the constant
$x_{02}$, in fact, simply taking $x_{02}>0$ will suffice. Such a
singularity remains on the right of the defect as
$t\rightarrow\infty$.

 The KdV
equation also allows other progressing singular solutions of the following
kind,
\begin{equation}\label{KdVsingular}
 \fl   u=\frac{2}{(x-x_1-ct)^2}-\frac{c}{6}, \qquad p=-\frac{2}{x-x_1-ct}-\frac{c}{6}\left(
    x-\bar x_1-\frac{1}{2}ct\right),
\end{equation}
where $c, x_1, \bar x_1$ are constants. The jump-defect will also affect
these in the following manner. Let $q$ be given by
\begin{equation}\label{}
    q=-\frac{2}{x-x_2-ct}-\frac{c}{6}\left(
    x-\bar x_2-\frac{1}{2}ct\right),
\end{equation}
where $x_2, \bar x_2$ are two additional constants. Then the
defect conditions require
\begin{equation}\label{}
    \bar x_1-\bar x_2 =-\frac{12}{c(x_1-x_2)}, \qquad \alpha=-\left(\frac{c}{6}+
    \frac{1}{(x_1-x_2)^2}\right).
\end{equation}
Again, as $\alpha$ becomes large $x_1\rightarrow x_2$ and the effect of the
defect on the singular solutions $u$ and $v$ disappears. However,
the non-locality will require a large value of $\bar x_1-\bar
x_2$.

\section{Discussion}

The purpose of this investigation has been to discover to what
extent the properties of jump-defects, originally described in
certain relativistic integrable field theories, extend to
non-relativistic systems of various kinds. In all cases, the
defects are purely transmitting, in the sense that solitons
passing through them may be delayed, converted to anti-solitons in
some cases, or absorbed, but will never be reflected. In every
case examined, the integrable defect conditions investigated would
constitute a B\"acklund transformation except that the spatial
derivatives of the fields are frozen at the defect location. It is
known that B\"acklund transformations for a given model are not
unique and, particularly for the KdV and mKdV models, several
different expressions are available in the literature. As
B\"acklund transformations, all the available expressions are
equivalent; but, as defect conditions, this does not appear always
to be the case. For instance, for both the KdV and the mKdV
equations local expressions, using $u$ instead of $p=u_x$, for a
B\"acklund transformation are available \cite{Chen74}. However, it
does not seem that all formulations of B\"acklund transformations
may be used as defect conditions since it is not always possible
to give a Lagrangian description for the $u$-expressions for these
models, and consequently to find suitable defect potentials.
Moreover, to allow momentum conservation, which is a key feature
of the jump-defect, the corresponding defect conditions for KdV
and mKdV need to satisfy the additional relation listed at the end
of \sref{section10}. Not all B\"acklund transformations seem to
satisfy such a condition - for an example of one that does not,
consider the B\"acklund transformation proposed in \cite{D&J} for
the KdV equation.

It ought to be possible to modify appropriate Lax pairs
for KdV and mKdV \cite{Ablowitz73}  to accommodate the jump-defects.
However, because of the proliferation of higher derivatives in the
expressions for the B\"acklund transformations this appears to
be less than straightforward and discussions of these will be deferred.

The nonlinear Schr\"odinger model seems to provide the nicest
results. In this case, bound states have been found and it has
been demonstrated using the two-soliton example that several
solitons passing a defect will be affected by it independently of
one another. If there are several jump-defects at different
locations they too will act independently. It transpires that the
strangest situation is illustrated by the KdV example. First, it
does not seem possible to recover the single bulk soliton by
taking a suitable limit of the defect parameter, and second the
consequences of taking the limit $|\beta|\rightarrow 0$ remains
mysterious. In fact, the situation in which $|\beta|<a$, that is
the defect parameter is less than the soliton speed, does not seem
to be allowed since the incoming soliton at the left of the defect
would become singular on the right of it. One resolution of this
that has been suggested is that the soliton is trapped by the
defect (as is clearly the case when $|\beta|=a$).

It was noted in \cite{Bow05} that the jump-defects of the
sine-Gordon model may themselves move and indeed scatter if there
are several moving with different speeds. This aspect has not been
considered in the present study though similar properties are
expected, at least for NLS. The non-local nature of the defect
conditions for the other models (KdV and mKdV) may pose a problem
in this regard. It was also noted in \cite{Bow05} that the natural
solution of the `triangle relations' \cite{Konik97},
\begin{equation}\label{STT}
 \fl   S_{kl}^{mn}(\Theta)\, ^{\rm e}T_{n\alpha}^{t\beta}(\theta_1)\,
    ^{\rm e}T_{m\beta}^{s\gamma}(\theta_2)
    =\, ^{\rm e}T_{l\alpha}^{n\beta}(\theta_2)\, ^{\rm e}T_{k\beta}^{m\gamma}(\theta_1)\,
    S_{mn}^{st}(\Theta)\, ,\qquad \Theta=\theta_1-\theta_2,
\end{equation}
where the roman labels are $\pm 1$, corresponding to soliton or
anti-soliton, the greek labels are even integers, corresponding to
the `topological charge' carried by the defect,
$S_{kl}^{mn}(\Theta)$ is the bulk scattering matrix for a pair of
solitons, and $^{\rm e}T_{n\alpha}^{t\beta}(\theta)$ is the
transmission matrix for a soliton passing through an even-charged
defect, is consistent with the Lagrangian description of the
sine-Gordon jump-defect. In the present context, the NLS model
appears to be the most suitable for a quantum investigation and
also the most interesting due to the presence of bound states. It
is well known that the quantum version of NLS is equivalent to
a one dimensional multi-particle problem with
$\delta$-function pairwise interactions of equal strength. The
problem in the bulk has been studied long ago, in the first
place by means of the coordinate Bethe ansatz \cite{Lieb63} (but see
also \cite{Korepin93} for a review and more references). It is
therefore natural to try to incorporate the jump-defect within the
$N$ body quantum picture and to discover what manner of particle
interaction is able to describe it. More recently,  an $N$ body
problem of this kind with impurity (in a `repulsive' regime where
there would be no solitons in the corresponding NLS model),
 has been analysed in
\cite{Crampe05}. However, more investigations are necessary to see
if there are any connections between that type of impurity and the
jump-defect described in this article.

One aspect of the story that remains frustrating is the absence of
a physical model of an integrable jump-defect. It would be very
interesting to discover a physical situation where the B\"acklund
transformation plays a natural role. There are jump-defects
commonly occurring naturally - modelling a dislocation in a
material provides an example, as does modelling a shock front, or
a fluid bore - where certain physical quantities are regarded as
discontinuous, such as the fluid velocity on either side of the
bore, yet others are continuous; and some conservation laws are
preserved. However, there does not yet appear to be a nonlinear,
and integrable, example of the kind we are discussing. If there
were,
 it might offer the possibility of controlling
solitons, which might then be put to use (see
\cite{cz} for a suggestion).

\ack
CZ thanks the Japan Society for the Promotion of Science for
a Fellowship and EC is indebted to members of the \'Ecole Normale
Sup\'erieure de Lyon, the University of Bologna and the Yukawa
Institute for Theoretical Physics, especially Jean Michel Maillet,
Francesco Ravanini and Ryu Sasaki, for their hospitality. Both
authors also wish to thank Davide Fioravanti and Peter Bowcock for
discussions on associated topics, and Ryu Sasaki for a critical reading of the
manuscript. The work has been supported in
part by the Leverhulme Trust and by EUCLID - a European Commission
RTN  (contract number HPRN-CT-2002-00325).

\end{document}